\documentclass{iau} 
\usepackage{graphicx,natbib}

{\newif\ifnotend
\notendtrue
\def\veclist{ABCDEFGHIJKLMNOPQRSTUVWXYZabcdefghijklmnopqrstuvwxyz.}
\def\top#1#2.{#1}
\def\tail#1#2.{#2.}
\loop\expandafter\xdef\csname v\expandafter\top\veclist\endcsname%
{{\noexpand\bf\expandafter\top\veclist}}
\edef\veclist{\expandafter\tail\veclist}
\if\veclist.\notendfalse\fi\ifnotend\repeat}
\def\spose#1{\hbox to 0pt{#1\hss}}
\def\lta{\mathrel{\spose{\lower 3pt\hbox{$\mathchar"218$}}
     \raise 2.0pt\hbox{$\mathchar"13C$}}}
\def\i{{\rm i}}

\title[Galaxy simulations] 
{Galaxy simulations in the Gaia era}

\author[Ivan Minchev]   
{Ivan Minchev}

\affiliation{Leibniz-Institut f\"{u}r Astrophysik Potsdam (AIP), An der Sternwarte 16, D-14482, Potsdam, Germany 
\\email: {\tt iminchev@aip.de}}

\pubyear{2017}
\volume{330}  
\jname{Astrometry and Astrophysics in the Gaia sky}
\editors{Alejandra Recio-Blanco, Patrick de Laverny, Anthony Brown and, Timo
Prusti eds.}
\begin{document}

\maketitle

\begin{abstract}
We live in an age where an enormous amount of astrometric, photometric, asteroseismic, and spectroscopic data of Milky Way stars are being acquired, many orders of magnitude larger than about a decade ago. Thanks to the Gaia astrometric mission and followup ground-based spectroscopic surveys in the next 5-10 years about 10-20 Million stars will have accurate 6D kinematics and chemical composition measurements. KEPLER-2, PLATO, and TESS will provide asteroseismic ages for a good fraction of those. In this article we outline some outstanding problems concerning the formation and evolution of the Milky Way and argue that, due to the complexity of physical processes involved in the formation of disk galaxies, numerical simulations in the cosmological context are needed for the interpretation of Milky Way observations. We also discuss in some detail the formation of the Milky Way thick disk, chemodynamical models, and the effects of radial migration.
\end{abstract}

\firstsection 
\section{Introduction}

The goal of Galactic Archeology \citep{freeman02} is to dissect the Milky Way into its various components (discs, bulge, bar and halo) and thus to disentangle the various processes that contributed to their formation and evolution. The importance of this topic is manifested in the number of Galactic surveys dedicated to obtaining spectroscopic information for a large number of stars, e.g., RAVE \citep{steinmetz06}, SEGUE \citep{yanny09}, APOGEE \citep{majewski10}, HERMES \citep{freeman10}, Gaia-ESO \citep{gilmore12}, and LAMOST \citep{zhao06}. This effort will soon be complemented by more than a billion stars observed by the Gaia space mission \citep{perryman01}. Millions of these will have accurate proper motions and parallaxes, which together with existing spectroscopic data, and especially with the advent of the dedicated Gaia follow-up ground-based surveys WEAVE \citep{dalton12} and 4MOST \citep{dejong12}, will enable Galactic Archaeology as never before.  

While in axisymmetric discs energy and angular momentum are conserved quantities and are, thus, integrals of motion \citep{bt08}, this is not true for the more realistic case of potentials including perturbations from a central bar and/or spiral arms. In the case of one periodic perturbation there is still a conserved quantity in the reference-frame rotating with the pattern -- the Jacobi integral $J=E-L\Omega_p$, where $E$ is the energy of the particle, $L$ is its angular momentum, and $\Omega_p$ is the pattern angular velocity. This is no longer the case, however, when a second perturbation with a different patterns speed is included. 

It has now been well established that the Milky Way disc contains both a bar (as in more than 50\% of external disc galaxies) and spiral structure moving at different pattern speeds, making it difficult to solve such a dynamical system analytically. Instead, different types of numerical methods are usually employed, from simple test-particle integrations, to preassembled N-body and SPH systems, to unconstrained, fully cosmological simulations of galaxy formation. All of these techniques have their strengths and weaknesses. Test particles are computationally cheap, allow for full control over the simulation parameters (such as spiral and bar amplitude, shape, orientation and pattern speed) but lack self-gravity. N-body simulations offer self-consistency but bar and especially spiral structure parameters are not easy to derive and not well controlled. Finally, in addition to being very computationally intensive, the outcomes of hydrodynamical cosmological simulations are even less predictable, with merging satellites and infalling gas making it yet harder to disentangle the disc dynamics; these are, however, much closer to reality in their complexity and a necessary ultimate step in the interpretation of observational data.

\subsection{Some of the questions we would like to answer}

\begin{itemize}
\item What are the spiral structure parameters?
\item Bar parameters?
\item Bulge structure?
\item Disk chemo-kinematical structure as a fn of radius and distance from plane?
\end{itemize}

The above four questions are concerned with the current disk state and need to be answered before we can proceed to the following ones:

\begin{itemize}
\item How did the Milky Way thick disk form?
\item How and when did the bulge/bar form?
\item How much radial mixing happened in the disk (fn of time and radius)?
\item Did the disk from from the inside-out?
\end{itemize}

For the latter set of questions a disk formation model covering the entire Milky Way history is needed. The complex dynamics of stars in the Galaxy demands the use of N-body simulations. Only then can we properly account for the perturbative effect of spiral arms, central bar, and minor mergers resulting from infalling satellites.

\section{The need for simulations in the cosmological context}

Here we present some of the complexity of physical processes encountered in the formation and evolution of galactic discs.

\subsection{Resonances in galactic discs}

Galactic discs rotate differentially with nearly flat rotation curves, i.e., constant circular velocity as a function of galactic radius. In contrast, density waves, such as a central bar and spiral structure, rotate as solid bodies. Therefore stars at different radii would experience different forcing due to the non-axisymmetric structure. Of particular interest are locations in the disc where the stars are in resonance with the perturber. The corotation resonance (CR), where stars move with the pattern, occurs when the angular rotation rate of stars equals that of the perturber. The Lindblad resonances (LRs) occur when the frequency at which a star feels the force due to a perturber coincides with the star's epicyclic frequency, $\kappa$. As one moves inward or outward from the CR circle, the relative frequency at which a star encounters the perturber increases. There are two values of $r$ for which this frequency is the same as the radial epicyclic frequency. This is where the inner and outer Lindblad resonances (ILR and OLR) are located. Quantitatively, LRs occur when the pattern speed $\Omega_{p}=\Omega\pm\kappa/m$, where $m$ is the multiplicity of the pattern\footnote{$m=2$ for a bar or a two-armed spiral structure and $m=4$ for a four-armed spiral.}. The negative sign corresponds to the ILR and the positive to the OLR. While Bertil Lindblad defined these for the case of an $m=2$ pattern (thus strictly speaking the ILR/OLR are the 2:1 resonances), for an $m=4$ pattern the ILR/OLR must be the 4:1 resonances.

\begin{figure*}
\centering
\includegraphics[width=\linewidth]{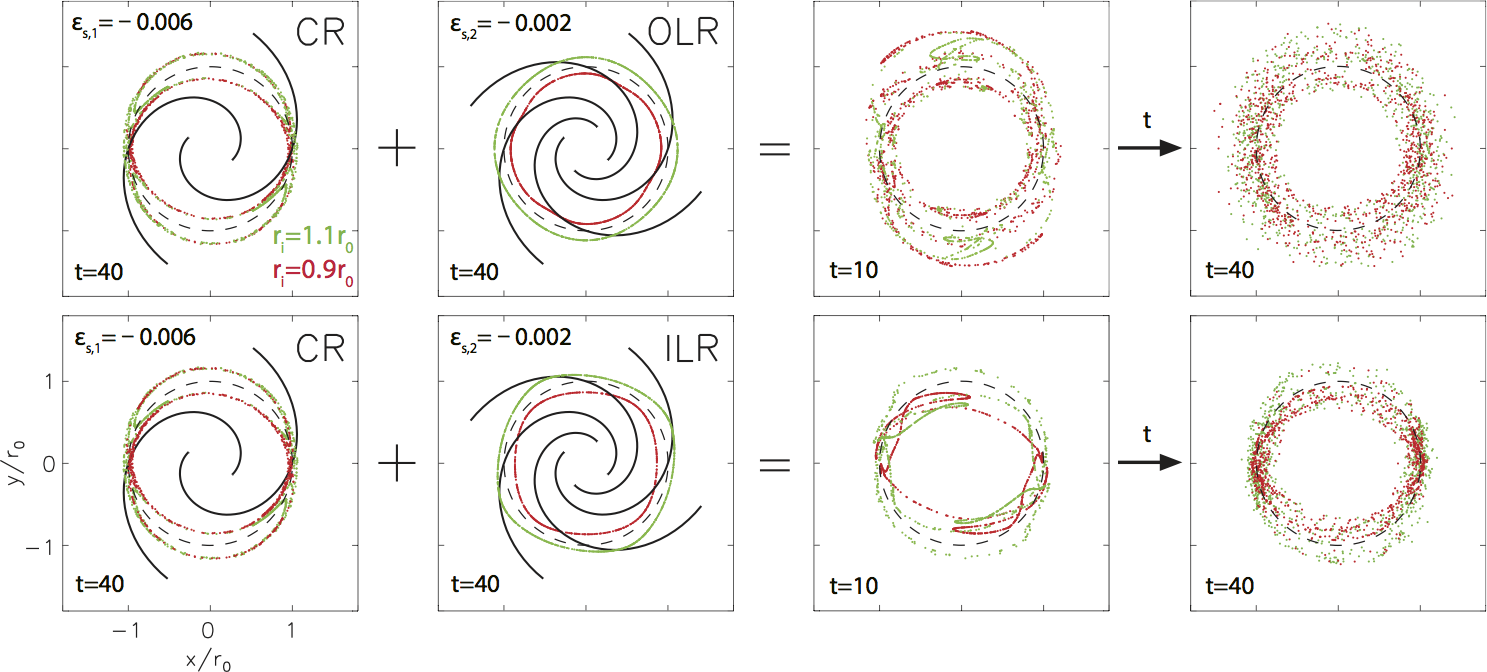}
\caption{ 
The effect on stellar orbits of two spiral perturbation with different pattern speeds acting together. {\bf First column:} The red/green particles initially start on rings just inside/outside the CR (dashed black circle) of a two-armed spiral wave. Distances are in units of the solar radius, $r_0$. The top and bottom plots are identical. {\bf Second column:} Same initial conditions, but near the 4:1 OLR (top) or 4:1 ILR (bottom) of a four-armed wave. {\bf Third column:} Same initial conditions, but particles are perturbed by both spiral waves shown at t=10 rotation periods, or about 2.5 Gyr. Note that a secondary wave of only 1/3 the strength of the first one is enough to disrupt the horseshoe orbits near the CR. {\bf Fourth column:} Same setup as in the third column but at t=40, or about 10 Gyr. Simulations from \cite{mq06}.
\label{fig:mq06}
}
\end{figure*}

\subsection{Multiple patterns in galactic discs}

According to the theoretical work by \cite{tagger87} and \cite{sygnet88}, two patterns can couple non-linearly as they overlap over a radial range, which coincides both with the CR of the inner one and the ILR of the outer one. This coincidence of resonances results in efficient exchange of energy and angular momentum between the two patterns. 

Figure \ref{fig:mq06} shows the effect of multiple patterns on a ring of stars, i.e., initially uniformly distributed in azimuth and at the same galactic radius. One can see that the orderly behavior under the influence of a single perturber becomes stochastic when two spiral waves moving at different patterns speeds are considered. 

\section{Effects of radial migration on disk thickening}

Several works have previously suggested that radial migration can give rise to thick disc formation by bringing out high-velocity-dispersion stellar populations from the inner disc and the bulge. Such a scenario was used, for example, in the analytical model of \cite{schonrich09b}, where the authors claimed to explain the Milky Way thick- and thin-disc characteristics (both chemical and kinematical) without the need of mergers or any discrete heating processes. Similarly, the increase of disc thickness with time found in the simulation by \cite{roskar08a} has been attributed to migration in the work by \cite{loebman11}. 

{\bf Migration induces disc flaring in isolated discs}: A first effort to demonstrate how exactly radial migration affects disc thickening in dynamical models was done by \cite{minchev12b}. It was shown that stellar samples arriving from the inner disc have slightly higher velocity dispersions, which will result in them being deposited at higher distances above the galactic midplane. However, the opposite effect arrises from samples arriving from the outer disc (with lower velocity dispersions). Therefore, the {\it overall} migration effect on the disc thickening is minimal throughout most of the disc extend, except in the very inner/outer parts of the disc, where only inward/outward migrators are deposited. This naturally results in disc flaring, as shown in Fig. 7 by \cite{minchev12b}. We explained this as the conservation of vertical action as opposed to conservation of the vertical energy assumed before. Several independent groups, using different simulation techniques and setups, have confirmed that migration does not thicken the disc (\citealt{martig14b, vera-ciro14, grand16}). 

{\bf Migration suppresses disc flaring when infalling satellites are present}: It is well known from both observations and cosmological simulations that minor mergers take place in the formation of galactic discs. Such interactions will have the effect of heating more the disc outskirts at any given time of the disc growth (more so at high redshift), because of the low mass density there. 

Interestingly, and to complicate matters, when this more realistic scenario is considered, migration has the opposite effect on the disc vertical profile compared to the effect of an isolated galaxy -- the role of outward and inward migrators is reversed in that they now cool and heat the disc, respectively. More on this can be found in \cite{mcm14} and the review by \cite{minchev17a}. 

\section{Formation of galactic thick discs by the flaring of mono-age populations}
\label{sec:thick}

Stellar disc density decomposition into thinner and thicker components in external edge-on galaxies find that thicker disc components have larger scale-lengths than the thin discs (e.g., \citealt{yoachim06, pohlen07}). While this is consistent with results for the Milky Way when similar morphologically (or structural) definition for the thick disc is used (e.g., \citealt{robin96, ojha01}), it is in contradiction with the more centrally concentrated older or [$\alpha$/Fe]-enhanced stellar populations (e.g., \citealt{cheng12b, bovy12a}). This apparent discrepancy may be related to the different definition of thick discs - morphological decomposition or separation in chemistry.

Additionally, while no flaring is observed in external edge-on discs \citep{vanderkruit82, comeron11}, numerical simulations suggest that flaring cannot be avoided due to a range of different dynamical effects. The largest source is most likely satellite-disc interactions (e.g., \citealt{villalobos08}), which have been found to increase an initially constant scale-height by up to a factor of $\sim10$ in 3-4 disc scale-lengths. Other sources of disc flaring include misaligned gas infall \citep{scannapieco09} and reorientation of the disc rotation axis. 

\begin{figure}
\centering
\includegraphics[width=14 cm]{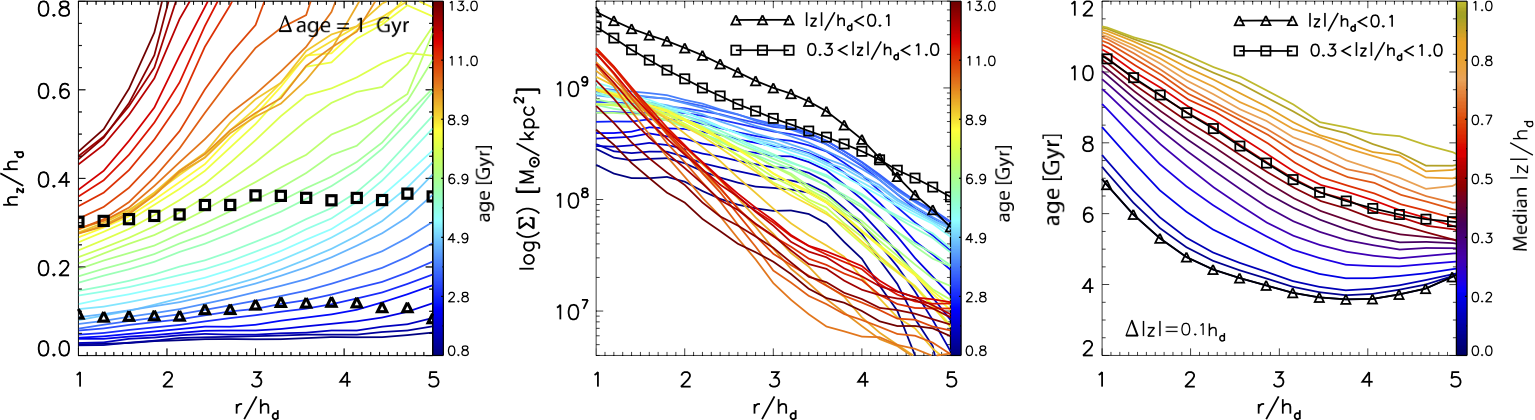}
\caption{
{\bf Left:} Variation of disc scale-height, $h_z$, with galactic radius for a cosmological disc formation simulation. Color lines show mono-age populations, as indicated. Overlapping bins of width $\Delta$age $=1$~Gyr are used. Overlaid also are the thin (triangles) and thick (squares) discs obtained by fitting a sum of two exponentials to stars of all ages. No significant flaring is found for the thin and thick discs. 
{\bf Middle:} Disc surface density radial profiles of mono-age populations. Older discs are more centrally concentrated, which explains why flaring diminishes in the total population. Also shown are the surface density profiles of stars close to (triangles) and high above (squares) the disc midplane. The thicker disc component extends farther out than the thin one, consistent with observations of external galaxies.
{\bf Right:} Variation of mean age with radius for samples at different distance from the disc midplane, as indicated. Slices in $|z|$ have thickness $\Delta|z|=0.1h_d$. Overlaid are also the age radial profiles of stars close to (triangles) and high above (squares) the disc midplane. Age gradients are predicted for both the (morphologically defined) thin and thick discs. Adapted from \cite{minchev15}.
\label{fig:thick1}
}
\end{figure}

\cite{minchev15} studied the formation of thick discs using two suites of simulations of galactic disc formation, one of which we present here. This is a full cosmological zoom-in hydro simulation, using initial conditions from one of the Aquarius Project haloes \citep{springel08, scannapieco09}. Further details about this simulation can be found in \cite{aumer13b}, their model Aq-D-5.

In the left panel of Fig.~\ref{fig:thick1} we plot the scale-height variation with galactocentric radius, $r$, in the region 1-5 disc scale-lengths, $h_d$. Both the radius and scale-height, $h_z$, are in units of $h_d$. It can be seen that significant flaring is present, which increases for older coeval populations. 

In contrast to the flaring found for all but the youngest mono-age populations, the thin and thick disc decomposition of the total stellar population including all ages, results in no apparent flaring. This is shown by the triangle and square symbols overlaid in the left panel of Fig.~\ref{fig:thick1}.

What is the reason for the flaring of mono-age discs? In numerical simulations flaring is expected to result from a number of mechanisms related to galactic evolution in a cosmological context (e.g., \citealt{kazantzidis08, villalobos08}). Even in the absence of environmental effects, flaring is unavoidable due to secular evolution alone (radial migration caused by spirals and/or a central bar, \citealt{minchev12b}). It should be stressed here that, while migration flares discs in the lack of external perturbations, during satellite-disc interactions it works {\it against} disc flaring (\citealt{mcm14}). Yet, this is not sufficient to completely suppress the flaring induced by orbiting satellites, as evident from the left panel of Fig.~\ref{fig:thick1}. This suggests that external effects are much more important for the disc flaring in this simulation. Because the mass and intensity of orbiting satellites generally decreases with decreasing redshift, so does the flaring induced. It can be expected that at a certain time secular evolution takes over the effect of external perturbations.\footnote{\cite{minchev14} suggested that the time at which internal evolution takes over can also be inferred from the shape of the [$\alpha$/Fe]-velocity dispersion relation of narrow metallicity samples.}

What is the reason for the lack of flaring in the total disc population? In an inside-out formation scenario, the outer disc edge, where flaring is induced, moves progressively from smaller to larger radii because of the continuous formation of new stars in disc subpopulations of increasing scale-length. At the same time the frequency and masses of perturbing satellites decreases. Because of the inside-out disc growth, which results in more centrally concentrated older samples (see Fig.~\ref{fig:thick1}, middle panel), the younger the stellar population, the further out it dominates in terms of stellar mass. The geometrically defined thick disc, therefore, results from the imbedded flares of different coeval populations, as seen in the left panel of Fig.~\ref{fig:thick1}. 

The right panel of Fig.~\ref{fig:thick1} shows that a geometrical thick disc is expected to have a negative age gradient. For this particular simulation the mean age decreases from $\sim10.5$ to $\sim6$~Gyr in four disc scale-lengths. Such an age drop of mean stellar age at high distances, $|z|$, from the disc midplane explains the inversion in $[\alpha$/Fe] gradients with increasing mean $z$ found by \cite{anders14} in APOGEE data.

The recent work by \cite{bovy16} estimated the variation of disk scale-height for mono-abundance populations (MAPs) of APOGEE red clump giants and found that the high-[$\alpha$/Fe] subpopulations did not show flaring, thus disagreeing with the scenario presented in this section. It was then shown by \cite{minchev17b} that flaring in MAPs can be lost if they are not mono-age populations as has been thus far assumed. This is because, due to the inside-out disk formation, a MAP displays a negative age gradient (except for the lowest [$\alpha$/Fe] MAPs). Flaring in the oldest APOGEE red clump stars was subsequently confirmed by \cite{mackereth17}, who used the same data and methods as Bovy et al. and ages estimated by \cite{martig16}. This example emphasizes the importance of good ages estimates for disentangling the structure and formation of the Milky Way.

\section{Chemo-dynamical modeling of the Milky Way}

So far we have focused on the dynamics of the Milky Way disc, which tells us mostly about its current state. To be able to go back in time and infer the Milky Way evolutionary history, however, we need to include both stellar chemical and age information. 

A major consideration in a disc chemo-dynamical model is taking into account the effect of radial migration, i.e., the fact that stars end up away from their birth places. Below we briefly summarize models which include radial migration. 
\newline$\bullet$ Semi-analytical models tuned to fit the local metallicity distribution, velocity dispersion, and chemical gradients, etc., today (e.g., \citealt{schonrich09a, kubryk15}) or Extended distribution functions \citep{sanders15}: 
\newline$-$ Easy to vary parameters
\newline$-$ Provide good description of the disc chemo-kinematic state today
\newline$-$ Typically not concerned with the Milky Way past history
\newline$-$ Time and spatial variations of migration efficiency due to dynamics resulting from non-axisymmetric disc structure is hard to take into account.
\newline$\bullet$ Fully self-consistent cosmological simulations (e.g., Kawata and Gibson 2003; Scannapieco et al. 2005; Kobayashi and Nakasato 2011; Brook et al. 2012):
\newline$-$ Dynamics self-consistent in a cosmological context
\newline$-$ Can learn about disc formation and evolution
\newline$-$ Not much control over final chemo-kinematic state
\newline$-$ Problems with SFH and chemical enrichment due to unknown subgrid physics
\newline$-$ Much larger computational times needed if chemical enrichment included.
\newline$\bullet$ Hybrid technique using simulation in a cosmological context + a classical (semi-analytical) chemical evolution model (\citealt{mcm13}:
\newline$-$ Avoids problems with SFH and chemical enrichment in fully self-consistent models
\newline$-$ Can learn about disc formation and evolution
\newline$-$ Not easy to get Milky Way-like final states.

\begin{figure}
\centering
\includegraphics[width=12.cm]{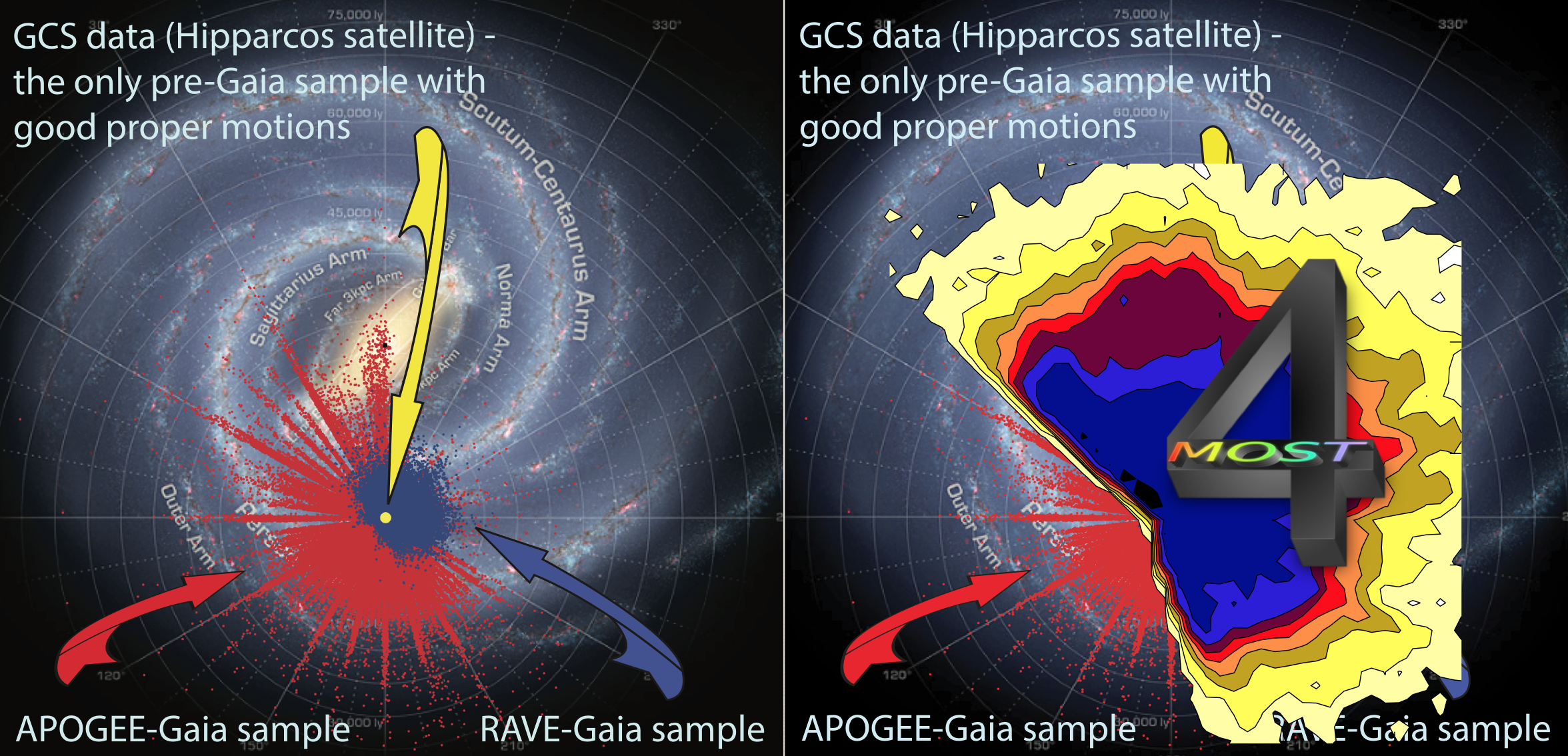}
\caption{
{\bf Left}: Illustrating the vast increase of stars with precise 6D kinematical information and chemical abundances after Gaia's first two data releases. Distribution of RAVE-Gaia and APOGEE-Gaia stellar samples overlaid on top of R. Hurt's map of the Milky Way (SSC-Caltech). The Galactic disc rotation is in the clockwise direction. 
{\bf Right}: The further increase expected from the ESO funded 4MOST spectroscopic survey, starting operation in 2022, that will provide radial velocities and chemistry for $10-15\times10^6$ Gaia stars.
\label{fig:mw}
}
\end{figure}

\section{Conclusions}

In this work we argued that numerical simulations of disk formation in the cosmological context are invaluable for the interpretation of the massive new data sets expected in the very near future. Here are some examples of outstanding problems that we can hope to be able to answer in the next 5-10 years:

\begin{itemize}
\item What is the pattern speed and length of the Galactic bar: fast bar, about 3 kpc long (Dehnen 2000, Fux et al. 2001, Minchev et al. 2007, 2010, Antoja et al. 2008, Monari et al. 2017) or Slow bar, about 5 kpc long (Wegg et al. 2015, Perez-Villegas 2017)?
 \item What is the nature of the Hercules stream: results from the bar's Outer Lindblad Resonance or from the bar's Corotation?
\item What causes the vertical wave patterns seen in the Milky Way disk: caused by bar/spirals or by the tidal effect of the Sgr dwarf galaxy?
\item Are the local vertical disk asymmetries and Monoceros Ring part of the same global structure?
\item How did the bulge form: inward stellar migration gives rise to different stellar populations in the bulge \citep{dimatteo16} or bulge chemistry distinct from that of the thick disk (Johnson et al. 2014)?
\item What is the shape of the age-velocity relation: increase of stellar velocity dispersion with age as a power law or a step up exists at 8-10 Gyr due to the heating effect of the last massive merger?
\end{itemize}

The field of Galactic Archaeology will soon be transformed by the Gaia end-of-mission astrometry and followup ground-based spectroscopic surveys such as RAVE, APOGEE, LAMOST, GALAH, WEAVE, and 4MOST, which in the next 5-10 years will deliver 10-20 Million stars with accurate 6D kinematics and chemical abundance measurements. Finally, we would like to emphasize the importance of good age estimates needed to break degeneracies and refine chemo-dynamical models, which we expect to get for a good fraction of the data from the asteroseismic missions KEPLER-2, PLATO, and TESS. 

\section{Acknowledgements}
IM acknowledges support by the Deutsche Forschungsgemeinschaft under the grant MI 2009/1-1.

{}

\end{document}